\documentclass[10pt,conference]{IEEEtran}
\IEEEoverridecommandlockouts

\usepackage{cite}
\usepackage{amsmath,amssymb,amsfonts}
\usepackage{algorithmic}
\usepackage{graphicx}
\usepackage{textcomp}
\usepackage{xcolor}
\usepackage{listings}
\usepackage{xspace}
\usepackage{ifthen}
\usepackage{url}
\usepackage{hyperref}
\usepackage{booktabs}
\def\BibTeX{{\rm B\kern-.05em{\sc i\kern-.025em b}\kern-.08em
    T\kern-.1667em\lower.7ex\hbox{E}\kern-.125emX}}

\begin{document}

\lstset{
    language=XML,
    morekeywords={build,plugins,plugin,dependency,groupId,artifactId,version,scope,dependencies,exclusions,exclusion},
    basicstyle=\footnotesize\ttfamily,
    captionpos=b,
    columns=fixed,
    keepspaces=true,
    frame=single,
    breaklines=true,
    postbreak=\raisebox{0ex}[0ex][0ex]{\ensuremath{\hookrightarrow\space}},
    showstringspaces=false,
    escapeinside={(*@}{@*)},
}

\newcommand{\etal}{\textit{et~al.}\xspace}
\newcommand{\ie}{\textit{i.e.,}\xspace}
\newcommand{\eg}{\textit{e.g.}\xspace}

\newcommand{\figref}[1]{Figure~\ref{#1}\xspace}
\newcommand{\chapref}[1]{Chapter~\ref{#1}\xspace}
\newcommand{\defref}[1]{Definition~\ref{#1}\xspace}
\newcommand{\lstref}[1]{Listing~\ref{#1}\xspace}
\newcommand{\secref}[1]{Section~\ref{#1}\xspace}
\newcommand{\tabref}[1]{Table~\ref{#1}\xspace}

\newcommand{\katIA}{katIA\xspace}
\newcommand{\katIALLM}{katIA-SP\xspace}
\newcommand{\katIAMA}{katIA-MA\xspace}
\newcommand{\dataset}{\textsc{Constantinople}\xspace}

\newcommand{\RQOne}{What is the performance of \katIALLM using static analysis and a single prompt to analyze the code change impact?}
\newcommand{\RQTwo}{To what extent does katIA's multi-agent variant (\katIAMA) improve change impact analysis compared to \katIALLM?}

\title{Hybrid Source Code Change Impact Analysis

\thanks{This research was funded in whole or in part by the Austrian Science Fund (FWF) 10.55776/P36698. For open access
purposes, the author has applied a CC BY public copyright license to any author accepted manuscript version arising
from this submission. The research was supported by the Austrian ministries BMIMI, BMWET and the State of Upper
Austria in the frame of the SCCH COMET competence center INTEGRATE (FFG 892418).}}

\author{
\IEEEauthorblockN{Katharina Stengg, Christian Macho, and Martin Pinzger}
\IEEEauthorblockA{
University of Klagenfurt\\
Klagenfurt, Austria\\
\{katharina.stengg, christian.macho, martin.pinzger\}@aau.at
}
}

\maketitle

\begin{abstract}
Whenever software developers modify a piece of code, they need to understand the impact of their changes on other code entities, for instance, to find out which other code entities need to be updated, too.
Developers often use change impact analysis (CIA) in such situations.
While several approaches for CIA were proposed, they still struggle with either low precision or low recall, thereby increasing manual effort for developers to verify the results.
Furthermore, limitations in the used datasets and experimental setups can bias the results reported by many CIA approaches.

In this paper, we first address the dataset limitations by manually labeling 1,931 changed code entities stemming from 263 commits
of 22 open-source Java projects.
Next, we introduce two variants of \katIA, our code change impact analysis approach that combines static impact analysis with the reasoning capabilities of large language models (LLMs).
The first variant uses a single prompt (\katIALLM) for refining an initial static impact set while the second variant uses three specialized agents (\katIAMA).

The evaluation with our curated dataset shows that \katIALLM with Gemini 2.5 Flash achieved an F1-score of 22.4\% when matching impacted entities by their fully qualified names (FQN) and 27.0\% when matching them by their class and entity names (CE).
The latter shows an improvement of 2.0 percentage points over the state-of-the-art.
This score is further improved by \katIAMA which, using Gemini 2.5 Flash, yields an FQN-matching F1-score of 29.1\%
and a CE-matching F1-score of 33.3\%.
While these improvements come at the cost of increased median generation time from 0.570~s to 11.214~s and median token usage from 1,235 to 6,462 tokens,
the improved performance justifies the additional cost.
The comparison of three LLMs using \katIALLM finds that Gemini 2.5 Flash is the most consistent model.
These results indicate that combining static impact analysis with LLM-based refinement, particularly with the multi-agent variant, can improve code change impact analysis.
\end{abstract}


\section{Introduction}
\label{sec:intro}
Software systems continuously evolve through changes to their source code~\cite{lehman1980programs}.
Whenever software developers modify a piece of source code, they need to understand the impact of these changes on other parts of the system to ensure code functionality and quality. 
For example, changing a method's sorting criterion may affect callers that rely on the previous ordering of the returned data, requiring updates to preserve the behavior.

According to Lehnert~\cite{lehnert2011taxonomy}, developers face two fundamental questions when changing software: (1) which software entities need to be modified to implement the intended change (seed), and (2) which additional entities may be affected as a consequence of this modification (impacted).
Change impact analysis~(CIA) aims to identify such affected entities of a software system~\cite{lehnert2011taxonomy}.
In practice, change impact analysis can support software maintenance tasks, such as bug fixing and new feature development.
Kretsou \etal~\cite{kretsou2021change} found that developers benefit from knowledge about co-changing code parts, because identifying affected code parts often requires more effort than implementing the intended change itself.
However, understanding the full impact of code changes still comes with a high workload and insufficient tool support, particularly during code review~\cite{bacchelli2013expectations}.
Furthermore, Yadavally and Nguyen~\cite{Seed2Scope2026Yadavally} found that existing change impact analysis approaches often struggle with low precision or low recall, especially because not all code changes and their impacts are structurally connected.

Existing CIA approaches rely on different sources of information, including syntactic dependency analysis of source code changes~\cite{ren2004chianti},
co-evolution information~\cite{zhou2024enhancing}, and, more recently, transformer-based code representations~\cite{EnhancingATHENA2024Yan}.
Motivated by the improvements achieved with large language models (LLMs) in software engineering~\cite{khare2025understanding, fruntke2025automatically, xia2024automated, yang2026roadmap, Macho2026Agent-BasedDepRepair},
research has begun to investigate whether code change impact analysis can benefit from
them, too.
For example, Yadavally and Nguyen~\cite{Seed2Scope2026Yadavally} presented \textsc{Ripple}, a two-phase LLM-based approach
to predict the change impact from the changes in one method and the corresponding Jira bug report.

While \textsc{Ripple} is a promising approach, several aspects of its methodology and evaluation can be improved. 
First, for determining the main change, which Yadavally and Nguyen~\cite{Seed2Scope2026Yadavally} refer to as the seed edit location, they chose the structurally most central method
as the seed change, which is an overly simplified heuristic.
Other approaches, such as \textsc{Athena}~\cite{EnhancingATHENA2024Yan}, treat each changed method once as the seed change and the remaining changed methods as impacted entities.
We argue that such heuristics might bias the evaluation of CIA approaches.
Second, the dataset \textsc{Alexandria}~\cite{EnhancingATHENA2024Yan} used by both \textsc{Ripple} and \textsc{Athena} does not cover added methods, changed fields and classes,
and it does not account for language-specific features, such as method overloading in Java. For instance, \textsc{Ripple} matches impacted methods only by the source file name
and the simple method name without parameters.
To address these issues, we propose the dataset \dataset,
which extends \textsc{Alexandria} with field and class changes.
\dataset comprises a total of 1,931 changed code entities.
    Each entity is represented by its fully qualified name and manually labeled as either seed, impacted, or non-related.

Next, we introduce \katIA, a hybrid approach that first uses static code analysis
to provide an initial static impact set which \katIA then refines using LLMs.
For the LLM-based refinement, we
explore two variants, first the refinement with a single prompt called \katIALLM and second, the refinement with three
specialized agents, called \katIAMA.
A preliminary version of \dataset and initial results for \katIA are presented in an extended abstract~\cite{katIAExtendedAbstract2026Stengg},
whereas this paper presents dataset construction, both variants of \katIA, and the evaluation in detail.

Compared to \textsc{Ripple}, \katIALLM and \katIAMA both receive the seed changes, the initial static impact set, and the bug report, whereas the change planner in \textsc{Ripple}
receives only the bug report and the seed changes, and the reasoner receives the change intent, change plan and the candidate set within the dependence clusters (\ie initial impact set).
Furthermore, \katIA does not rely on evolutionary coupling but focuses on the static code relationships of the seed changes.
Finally, we do not restrict \katIALLM and \katIAMA to only refine the impact set within the clusters computed from the candidate set, but allow them to
extend the initial impact set. 

We used \dataset to evaluate the two variants of \katIA.
First, we studied the performance and consistency of \katIALLM to answer \textbf{RQ1: \RQOne}
We experimented with three LLMs and measured the performance
with precision, recall, and F1-score. To compare with the state-of-the-art, we computed these metrics
using CE-matching, based on class and entity names, and additionally using FQN-matching, based on fully qualified names.
Furthermore, we repeated each experiment three times to measure the consistency of results.
When using CE-matching, \katIALLM achieved F1-scores of 27.0\% with Gemini 2.5 Flash and 28.5\% with Qwen3 Coder Plus.
Gemini 2.5 Flash was the most consistent model, producing 99.62\% consistent results.
Furthermore, \katIALLM with Gemini 2.5 Flash and Qwen3 Coder Plus achieved F1-score improvements of 2.0 and 3.5 percentage points over the
F1-score reported for \textsc{Ripple}~\cite{Seed2Scope2026Yadavally}.

Second, we investigated whether decomposing the refinement task across three specialized agents
improves \katIALLM to answer \textbf{RQ2: \RQTwo}

For this experiment, we used only Gemini 2.5 Flash because of its high consistency and decent performance.
The results show a further increase in precision, recall, and F1-score under FQN-matching, the latter
reaching 29.1\%.
Under CE-matching, recall and F1-score increased, the latter reaching 33.3\%,
while precision decreased slightly.
Compared to \textsc{Ripple}, the CE-matching F1-score is 8.3 percentage points higher.

Overall, our results show that combining static code analysis with LLM-based refinement, especially in a multi-agent variant,
can improve code change impact analysis.
However, despite these improvements, change impact analysis remains a challenging problem.

In summary, this paper makes the following contributions:\\
\begin{enumerate}
    \item An extended, more detailed version of the \textsc{Alexandria} dataset, named \dataset, for evaluating code change impact analysis approaches, comprising manually labeled method, field, and class changes.
    \item A detailed analysis of the manual labeling process.
    \item \katIA, our hybrid code change impact analysis approach that combines static code analysis with LLM-based refinement, considering a single prompt (\katIALLM) and a multi-agent variant (\katIAMA).
    \item An evaluation of both variants and comparison to the state-of-the-art approach \textsc{Ripple}.
\end{enumerate}

The remainder of the paper is structured as follows:
\secref{sec:rel_work} discusses scientific work.
\secref{sec:dataset} explains our dataset labeling process and presents the descriptive statistics of \dataset.
\secref{sec:approach} presents our approach, including static code change impact set construction and LLM-based refinement.
\secref{sec:results} reports our experimental setup and results,
\secref{sec:discussion} discusses the results and threats to validity, and
\secref{sec:conclusion} concludes the paper.
The supplementary material of this work can be found online\footnote{\url{https://doi.org/10.6084/m9.figshare.32924534}}.
\section{Related Work}
\label{sec:rel_work}
This section discusses related work on change impact analysis and points out the research gaps addressed in this paper.

\subsection{Change Impact Analysis}
Several approaches to support developers in analyzing the impact of code changes were proposed.
Ren \etal~\cite{ren2004chianti} designed and implemented \textsc{Chianti}, a change impact analysis tool for Java.
\textsc{Chianti} receives two application versions and works on the atomic changes within them.
The resulting change impact is the set of tests potentially affected by the changes performed.
Furthermore, the authors reported the affecting changes, \ie the changes that could alter test execution.
Their empirical evaluation revealed that an average of 52\% of the unit tests of Daikon were affected.

Lehnert~\cite{lehnert2011taxonomy} proposed a taxonomy of impact analysis to help practitioners navigate and filter the variety of existing impact analysis approaches.
The taxonomy comprises eight classification criteria, including scope, granularity, analysis style, and the available experimental results.
Gethers \etal~\cite{gethers2012integrated} proposed an approach for change request impact analysis
that integrates history-based mining with dynamic analysis and information retrieval.
They evaluated their approach on four open-source projects, yielding statistically significant improvements
over the information retrieval-based impact analysis approach (IRCR). 
Dit \etal~\cite{dit2014impactminer} presented \textsc{ImpactMiner}, which implements
the approach proposed by Gethers \etal~\cite{gethers2012integrated}.

Hanam \etal~\cite{hanam2019aiding} addressed the problem of
code change understanding in the context of code review by proposing semantic change impact relations
and their tool \textsc{Semcia} which was evaluated on 2,000 JavaScript commits from three Node.js projects.
They were able to decrease the number of false positives and the size of change impact sets.
Finally, they conducted a user study where they found that their tool positively impacted both the speed and accuracy of the developers' code review tasks.
Huang~\etal~\cite{huang2021change} proposed an approach to improve traditional change impact analysis
by mapping historical change patterns from similar classes in other projects to boost the prediction accuracy of existing CIA tools.
Zhou~\etal~\cite{zhou2024enhancing} used evolutionary coupling to improve impact predictions.
Evolutionary coupling refers to relationships where code entities tend to frequently change together, similarly to the co-evolution of production and test code.
The authors discussed twelve types of software relationships that define the edges between code entities and can be used to model impact dependencies.

Yan \etal~\cite{EnhancingATHENA2024Yan} proposed \textsc{Athena}, a transformer-based ranked impact set prediction approach
using program dependence graphs.
They evaluated \textsc{Athena} using \textsc{Alexandria}, a dataset based on the set of untangled commits from Herbold~\etal~\cite{Untangling2022Herbold}.
As mentioned before, the dataset contains no clear indication of the seed change.
In their evaluation, each changed method
of a commit was treated as a query method, \ie seed change, once, while the other methods formed the corresponding ground-truth impact set.
This led to 4,405 impact analysis tasks in the largest of the three different evaluation settings, which were evaluated using ranked set metrics.

Yadavally and Nguyen~\cite{Seed2Scope2026Yadavally} proposed \textsc{Ripple}, an intent-aware, two-phase LLM-based approach to predicting impact sets from seed method changes and Jira bug reports.
For the evaluation, they created a dataset comprising the changed methods of 100 commits from the \textsc{Alexandria} dataset represented by their source-file and method names.
For each commit, they chose the method with the most direct and indirect dependencies, including call and class-member dependencies, as the seed change for their impact analysis task.
Using GPT-4o, \textsc{Ripple} outperformed \textsc{Athena} achieving an average precision of 28.2\%, recall of 36.3\%, and F1-score of 25.0\% for matching the source file name and the simple name of the method.
Finally, they report the results from an ablation study, demonstrating the importance of incorporating both the change intent and the LLM planning phase into their method.

Addressing the shortcomings of simple heuristics for determining the seed change, Stengg \etal~\cite{ERA2025Stengg} proposed a dataset extension of \textsc{Alexandria}, for which
they labeled the seed-impact relationships of changed code entities.
They evaluated the capabilities of the models GPT-5 and GPT-5-mini for CIA, obtaining median F1-scores of 0.0\%, which shows the limitations of approaches that rely solely on simple prompts and LLMs.

\subsection{Research Gap}
As mentioned before, existing CIA approaches, such as \textsc{Ripple}~\cite{Seed2Scope2026Yadavally} or \textsc{Athena}~\cite{EnhancingATHENA2024Yan},
rely on datasets, such as \textsc{Alexandria}, that have several limitations.
First, they consist of Java method changes only and ignore new methods and changes of other code entities such as fields and classes.
Second, they use simplified heuristics to determine the seed change~\cite{Seed2Scope2026Yadavally} or use each method as the seed change once~\cite{EnhancingATHENA2024Yan}.
Third, they do not fully support method overloading.
We argue that all three shortcomings can notably bias the results.
In our dataset \dataset presented in the next section, we address these shortcomings and lay the foundation for a less biased evaluation of change impact analysis approaches.

Furthermore, we propose \katIA, a hybrid approach combining static code analysis and LLMs.
Similar to \textsc{Ripple}, \katIA begins with a static analysis phase to produce an initial candidate set.
For \textsc{Ripple}, this set restricts LLM reasoning to a fixed dependency-enhanced candidate set, while \katIA treats the initial static impact set as a starting point and uses the LLM to 
both filter existing entities and identify additional impacted entities.
Furthermore, \katIA relies solely on static code relationships, eliminating the need for evolutionary coupling and project history.
We evaluate two variants of katIA, \katIALLM, which uses a single prompt for impact set refinement, and \katIAMA, which uses three agents for impact set refinement.
\section{Code Change Impact Dataset}
\label{sec:dataset}

This section presents the motivation, construction, and descriptive statistics of our dataset \dataset.
It extends \textsc{Alexandria}, a benchmark for impact analysis published by Yan \etal~\cite{EnhancingATHENA2024Yan}.
\textsc{Alexandria} is based on a set of manually untangled commits proposed by Herbold \etal~\cite{Untangling2022Herbold}. 
Yan~\etal state that there are other datasets using manually untangled commits.
However, these datasets either contain a limited number of commits or perform untangling at the file or commit level rather than the method level, which limits the applicability for method-level impact analysis.
The \textsc{Alexandria} dataset contains 910 commits from 25 open-source Java projects, which is
substantially more than other benchmarks so far.

\subsection{Limitations of \textsc{Alexandria}}
\label{subsec:shortcomings-of-textsc{alexandria}}
Despite being used for the evaluation of various state-of-the-art CIA approaches such as \textsc{Athena}~\cite{EnhancingATHENA2024Yan} and \textsc{Ripple}~\cite{Seed2Scope2026Yadavally},
we found that \textsc{Alexandria} has four main limitations, namely
    (1) limited entity coverage, (2) missing method signatures, (3) missing newly added methods, and (4) undefined seed changes.

\textbf{Limited entity coverage.} It only covers method changes but ignores changes to other program entities, such as fields and classes.
This limits the practical usability of the approaches as developers not only change methods.

\textbf{Missing method signatures. } Method changes are represented by the method names without their parameters, which
implies that approaches cannot precisely resolve overloaded methods.

\textbf{Missing added methods.} The dataset does not include newly added methods.
Consequently, the approaches using it are limited to methods that already exist in the corresponding parent commit.

\textbf{Undefined seed changes.}
Yan \etal~\cite{EnhancingATHENA2024Yan} state that the \textsc{Alexandria} dataset provides no clear indication of the seed change.
Instead, in their approach each changed method of a commit is treated as a seed change once, while the remaining methods form the corresponding impact set.
Yadavally and Nguyen~\cite{Seed2Scope2026Yadavally} also used \textsc{Alexandria} but chose a different approach
for determining the seed change.
They used the method with the most direct and indirect call and class-member dependencies as the seed change.
Both approaches solely rely on heuristics to identify the seed changes.

To address these limitations, we extended the \textsc{Alexandria} dataset in two ways.
First, in addition to method changes, we extracted changes to fields and classes, as well as newly added methods.
For each entity, we used its fully qualified name\footnote{\url{https://docs.oracle.com/javase/specs/jls/se11/html/jls-6.html\#jls-6.7}}, \eg\texttt{org.example.A.b()}, including parameter types for methods to handle method overloading.
Second, we manually labeled each changed code entity as either a seed, an impacted, or a non-related change.

\subsection{Dataset Construction}\label{subsec:dataset-labelling}
Change impact analysis (CIA) aims to identify the software entities impacted by a change~\cite{lehnert2011taxonomy}.
Accordingly, we annotated every changed production code entity either as seed, impacted, or non-related.

\textbf{Seed}. A seed change is a code change made to fulfill the requirements derived from the corresponding Jira bug report.
The set of all seeds for a bug report constitutes the bug-fixing change as defined by Rodríguez-Pérez \etal~\cite{rodriguez2020bugs}.
Unlike the heuristics used in \textsc{Ripple} or \textsc{Athena},
a single commit may contain multiple seeds when several changed entities are involved in addressing the reported bug.

\textbf{Impacted}. Impacted changes are triggered by the seed changes, not by the requirements of the reported bug.
They are consequences, \ie they do not directly address the bug report but become necessary because of the seed changes.
For example, if a seed change modifies a method's sorting criterion, its callers that rely on the previous ordering of the returned data must be modified to preserve the behavior. 

\textbf{Non-related}. Code changes that are neither seed nor impacted are non-related changes.
These changes occur within the same tangled commit~\cite{UntanglingCodeChanges2015Dias} but have no clear origin in or relationship to the bug report.
Although Herbold \etal~\cite{Untangling2022Herbold} untangled the dataset, we still found a portion of tangled changes during our dataset construction.

To classify changed code entities into these categories, we grouped the data of \textsc{Alexandria} by repository, commit hash, and parent commit hash.
This resulted in 918 commits.
In contrast, the 910 commits reported by the original authors correspond to grouping by repository and parent commit hash only.
We then extracted a list of changed files per commit and excluded test files based on standard directory and naming conventions.
Finally, we filtered the commits to retain only those that modified between one
and five Java source files for our dataset construction.
We selected this range because at least one changed file is required to contain changed entities, while an upper bound of five changed files keeps manual labeling feasible.
In the original dataset, only 94 of the 918 commits are not within this range.
Similar to Yadavally and Nguyen~\cite{Seed2Scope2026Yadavally}, we extended our dataset with the issue ID from the corresponding commit message to link the commits to
their Jira bug report.
The majority of the bug reports (82.87\%) refer to a single commit, while the maximum number of commits for a single bug report is eight. The latter occurs only once while 
111 bug reports (14.62\%) are referenced by exactly two commits, 13 by three, and five by four commits.
We excluded commits referencing the same bug report from the dataset only if they contained identical code changes or a subset of changes of
another commit in the dataset.
Changes to comments, imports, and annotations were excluded. 
Also note that commits which did not contain at least one seed change and one impacted change were skipped, adhering to the
criteria used in other works.

To label the changed code entities, we implemented a helper tool based on Spoon~\cite{Spoon2015Pawlak}
to extract the changed source code entities per commit.
Furthermore, its user interface helps evaluators to inspect the
changed code entities and the corresponding bug report. 
The labeled code entities are exported into a CSV file, where each row corresponds to a commit containing
metadata and one JSON object per change label with further information, such as the fully qualified name.
All evaluators used this tool for labeling. 

The labeling process was performed by the first author and two co-authors, all of them having several years of experience
in Java and code analysis. The labeling comprised three phases. 
We initially sampled 420 commits to use 20 commits for the calibration phase and the remaining 400 for the actual dataset labeling phases.
The calibration phase, where each evaluator labeled the same 20 commits, established a common understanding of the labeling guidelines and reduced ambiguities before the main labeling process.
The shared phase produced the first 100 labeled commits of the dataset, for which the main author calculated the agreement among the evaluators
to refine the guidelines and ensure consistent labeling.
Finally, the individual phase was used to extend the dataset, as each evaluator labeled 100 individual commits.

\subsubsection{Calibration Phase}
First, each evaluator received the same set of 20 randomly sampled commits to be labeled independently.
After that, they discussed their findings
to establish common labeling guidelines and practices to define a comprehensive labeling strategy for further labeling.

\subsubsection{Shared Phase}
In the shared labeling phase, each evaluator labeled the same 100 randomly sampled commits that were not part of the calibration set.
The first author collected the results from all evaluators and analyzed their pairwise overlap.
Majority vote was applied, and we discussed labels with strong disagreements. 
The first author also improved the guidelines with the learnings of this phase.

\begin{table}[htbp]
    \caption{Jaccard set similarities per evaluator pair and label of the shared phase.}
    \begin{center}
        \begin{tabular}{lrrrr}
            \toprule
            \textbf{Evaluator Pair} & \textbf{Seed} & \textbf{Impacted} & \textbf{Non-related} & \textbf{Overall} \\
            \midrule
            A--C & 50\% & 51\% & 64\% & 55\% \\
            A--B & 51\% & 55\% & 74\% & 60\% \\
            B--C & 48\% & 47\% & 62\% & 52\% \\
            \midrule
            Average & 50\% & 51\% & 67\% & 56\% \\
            \bottomrule
        \end{tabular}
    \end{center}
    \label{tab:jaccard_similarities}
\end{table}

Since each evaluator selected a label for each code entity,
we measured the pairwise agreement using Jaccard set overlap similar to prior work on agreement for set-valued annotations~\cite{passonneau2006measuring}.
We report the averages per evaluator pair and label in \tabref{tab:jaccard_similarities}.
During this shared phase, the overall set overlap was 56\%, with average pairwise overlaps ranging from 52\% to 60\%.
The similar values among the author pairs suggest that differences did not occur between a
specific evaluator pair. Furthermore, these values are not interpreted as inter-coder agreement, but as analysis of the set overlaps within the
shared labeling process.
An overlap of 100\% would mean that the sets reported by both authors were identical,
whereas missing or additional code entities per label category result in lower values.
We can observe that the agreement for non-related code changes is substantially higher than for seed or impacted changes.
This indicates that it is easier to choose changes that neither contribute to a bug fix nor result from it than to distinguish between
seed and impacted code entities.
The distinction between seed and impacted code entities is not always clear from the changes alone
but needs context and developer interpretation.

One evaluator documented all label discussions and the final decisions for previously undecided labels. 
This resulted in the first 100 annotated commits of our dataset, comprising 71 labeled and 29 skipped commits, as they did not contain at least one seed change and one impacted change and therefore were excluded from our dataset.
All evaluators agreed that previous discussions were valuable as preparation for the individual labeling phase
and helped resolve ambiguities as the distinction between seed and impacted entities may allow different interpretations.

\subsubsection{Individual Phase}
In the individual labeling phase, each evaluator labeled another set of 100 commits that were in neither of the two previous labeling phases.
The goal of this phase was to extend the dataset. 
\subsection{\dataset Dataset}\label{subsec:descriptive-dataset-statistics}

\tabref{tab:dataset} shows the descriptive statistics of the labeled dataset \dataset.
It comprises 263 commits and 1,931 changed code entities across 22 projects.
The number of changed code entities is computed as the sum of the entities labeled as seed changes (\#S),
impacted changes (\#I), and non-related changes (\#NR).
For each commit, the dataset also contains the commit hash of the current commit and its parent,
the number of changed production code files and code entities, the total number of added and deleted lines of code,
the commit message, the Jira bug report key, title, description, and the labeled entities represented by their fully qualified names.
\dataset is available in the supplementary material.

\begin{table}
    \centering
    \caption{Descriptive statistics of the labeled dataset \dataset. \#C refers to the number of commits,
        \#BR the number of bug reports in Jira, \#S, \#I, and \#NR list the number of changed code entities labeled as seed, impacted, and non-related,
    respectively.}
    \label{tab:dataset}
    \resizebox{\columnwidth}{!}{
    \begin{tabular}{lrrrrr}
        \toprule
        \textbf{Project} & \textbf{\#C} & \textbf{\#BR} & \textbf{\#S} & \textbf{\#I} & \textbf{\#NR} \\
        \midrule
        ant-ivy & 56 & 55 & 109 & 276 & 14 \\
        commons-bcel & 3 & 3 & 4 & 37 & 5 \\
        commons-beanutils & 5 & 5 & 9 & 16 & 1 \\
        commons-codec & 2 & 2 & 3 & 36 & 0 \\
        commons-compress & 19 & 14 & 37 & 109 & 2 \\
        commons-configuration & 21 & 20 & 32 & 69 & 5 \\
        commons-dbcp & 3 & 3 & 5 & 13 & 0 \\
        commons-digester & 4 & 3 & 6 & 15 & 0 \\
        commons-io & 9 & 9 & 12 & 44 & 3 \\
        commons-jcs & 10 & 10 & 17 & 36 & 1 \\
        commons-lang & 11 & 11 & 24 & 65 & 1 \\
        commons-math & 29 & 28 & 58 & 147 & 13 \\
        commons-net & 18 & 17 & 23 & 63 & 0 \\
        commons-scxml & 8 & 8 & 14 & 39 & 9 \\
        commons-validator & 5 & 4 & 22 & 31 & 0 \\
        commons-vfs & 10 & 10 & 30 & 58 & 6 \\
        deltaspike & 2 & 2 & 2 & 5 & 2 \\
        giraph & 18 & 18 & 40 & 92 & 5 \\
        gora & 5 & 5 & 11 & 17 & 33 \\
        opennlp & 12 & 12 & 23 & 45 & 0 \\
        parquet-mr & 12 & 12 & 16 & 109 & 6 \\
        systemml & 1 & 1 & 1 & 5 & 0 \\
        \midrule
        Total & 263 & 252 & 498 & 1327 & 106 \\
        \bottomrule
    \end{tabular}}
\end{table}

During the phases of dataset construction, the evaluators labeled a total of 420 randomly sampled commits.
The 20 samples used during the calibration phase were not considered for the final dataset.

Among the 400 commits, they identified 37 bug reports for which multiple commits referenced the same Jira bug report key.
For ten of these 37 bug reports, the corresponding commits contained no identical code changes,
indicating that they addressed different aspects of the same bug report.
For 20 bug reports, the commits contained identical code changes, while for seven bug reports, a later commit contained a subset of the code changes introduced by an earlier commit.
To avoid duplicates, we retained only the first occurrence of each such commit and excluded the remaining occurrences from the dataset,
resulting in 20 commits being removed from our final dataset.
As an additional consistency check, we compared the labels of these commits.
If commits contained a subset of the changes of another, we restricted this comparison to the shared code changes.
We found that the code changes for 18 bug reports received identical labels,
while four received partially identical labels with overlapping selected labels.
The commits of the remaining five bug reports were assigned different labels.

During the labeling process, the evaluators identified 117 commits that did not contain at least one seed change and one impacted change.
Similar to prior work that uses labeling heuristics to create sets containing at least one seed and one impacted entity, we excluded these 117 commits from the final version of \dataset.
Together with the previously discussed 20 commits that reference the same bug reports, 137 of the 400 commits were excluded from \dataset, resulting in 263 labeled commits that we use
for the evaluation of \katIA.
\section{katIA Approach}\label{sec:approach}
\begin{figure}[!t]
    \centering
    \includegraphics[width=0.9\columnwidth]{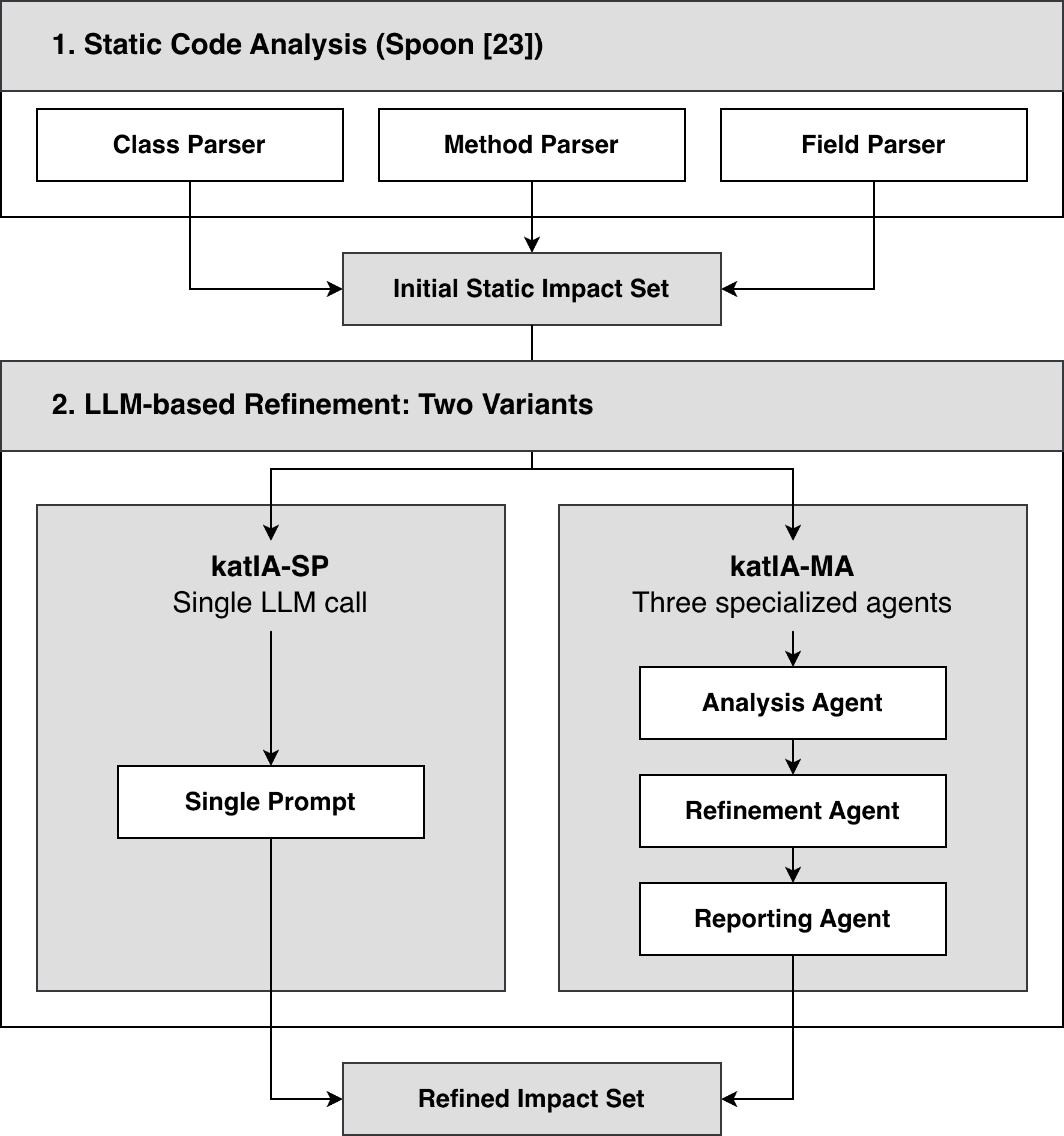}
    \caption{Overview of the \katIA approach that, given the seed changes of a commit, first constructs an initial static code change impact set that then is refined either with a single prompt (\katIALLM) or three agents (\katIAMA).}
    \label{fig:approach}
\end{figure}

\figref{fig:approach} provides an overview of our proposed approach \katIA.
\katIA consists of two phases.
In the first phase, given a set of seed changes, static analysis is used
to compute the initial static code change impact set.
This set consists of code entities connected to the seed changes through usage relationships.
For example, for a method, this set contains all its callers and callees.
In the second phase, the initial static code change impact set is refined by using LLMs.
The second phase exists in two variants, either using a single prompt (\katIALLM) or three agents (\katIAMA).
In the following, we describe each phase of \katIA in detail.

\subsection{Phase 1: Static Impact Set Construction}
\label{subsec:static-impact-set-construction}
To construct the initial static code change impact set, katIA first clones the corresponding project and checks out the respective parent commit.
Next, it uses the static code analysis framework Spoon~\cite{Spoon2015Pawlak} to parse the project and to identify the
corresponding code entity of each given seed change.
If a seed change is absent from the checked-out commit because the entity is newly added in the evaluated commit, the static analysis cannot resolve the corresponding code entity.
Consequently, the analysis returns an empty impact set for that entity.
Next, depending on the type of the code entity, katIA statically analyzes the
dependencies to other code entities.
katIA supports classes including interfaces, methods including constructors, initializer blocks, and fields.
Concerning fields, katIA extracts read, write, and read-write relations, \ie the static impact set for a field consists of a
list of methods that access it.
Concerning methods, \katIA extracts all callers and callees.
Classes can be used, extended, or implemented by other classes.
katIA extracts incoming and outgoing dependencies, as it identifies the classes that use, extend, or implement the seed class, and
the classes that the seed uses, extends, or implements.
The initial static impact set of the seed changes in a commit consists of all dependent code entities identified
by katIA. 
Currently, the static parser of katIA does not consider indirect dependencies between code entities, dynamic analysis during runtime, reflection-based accesses, or external libraries.
Consider an example where methods \emph{a()}, \emph{b()}, and \emph{c()} were changed, assuming that the seed change is in method \emph{a()}.
Method \emph{a()} calls method \emph{b()}, whereas \emph{c()} calls \emph{a()}.
The corresponding initial static code change impact set computed by katIA contains \texttt{org.example.A.b()} and \texttt{org.example.A.c()}.

\subsection{Phase 2: Impact Set Refinement}
\label{subsec:impact-set-refinement}
LLMs have been widely used to address software engineering challenges~\cite{sasaki2025landscape}.
Regarding code change impact analysis, a recent approach of Yadavally and Nguyen~\cite{Seed2Scope2026Yadavally} called \textsc{Ripple} showed
that a combination of dependency analysis in the first phase and LLMs in the second phase can improve the performance
of code change impact analysis. 
Motivated by \textsc{Ripple}, katIA also uses LLMs in the second phase to refine the initial static code change impact set computed by our static analysis.
We provide two variants of the second phase: first, using a single prompt (\katIALLM), and second, using three specialized agents (\katIAMA).

\subsubsection{Single Prompt Refinement}

We carefully constructed the prompt according to the findings and guidelines of recent software engineering research~\cite{wang2024advanced,ma2025should,fan2025exploring}.
The prompt template can be found in the supplementary material.
It starts with a system prompt that tells the LLM that it is an expert in Java and in analyzing the impact of source code changes on other code entities in a project's production code.
Such impersonation has been shown to enhance the performance of LLMs~\cite{salewski2023context}.
The prompt then presents general background information.
It states that the model receives information about a set of changes, called seed changes, that were made to fix a bug reported in Jira
and an initial static impact set containing the fully qualified names of code entities statically connected to the seed changes.
This general information is followed by the specific change impact analysis task.
This task tells the model that an impacted code entity must be created or modified as a consequence of the seed changes.
Additionally, this part states when to remove or add a code entity to the initial static impact set
and the final goal of the task, which is to refine the initial set to only contain impacted code entities.
Next, the prompt contains all inputs (Jira bug report, seed changes, and initial static code change impact set), followed by explicit instructions on the output format and three example outputs.

\subsubsection{Multi-Agent Refinement}

The goal of \katIAMA is to improve \katIALLM by decomposing the single prompt into three separate subtasks.
All agent and task definitions can be found in the supplementary material.
Each subtask corresponds to a separate state of the change impact analysis process and
is assigned to one specific agent.\\
\textbf{Analysis Agent. }
This agent is tasked to interpret the seed changes and the Jira bug report
to extract evidence relevant to identifying non-seed Java code entities
that likely need modification or creation as a consequence of the seed changes.
It produces evidence for each entity of the initial static code change impact set,
including their fully qualified name, an assessment of whether the entity should be modified as a consequence of the seed changes,
and the supporting evidence.
Similarly, it is tasked to produce this evidence for code entities that need to be created as a consequence of the seed changes.\\
\textbf{Refinement Agent. }
The structured evidence of the previous agent is provided to the refinement agent together with the seed changes and the initial static code change impact set.
Similar to the task description of \katIALLM, this agent is instructed to remove existing code entities from the
set and add missing ones when sufficient evidence was provided by the previous agent.\\
\textbf{Reporting Agent. }
The reporting agent is provided with the description of the expected output format and example outputs adhering to this format.
This agent is responsible for reporting the output in the expected format.
It is instructed to not alter the set received from the previous agent as its only responsibility is reporting the refined impact set in
the expected output format.
\section{Evaluation}
\label{sec:results}
This section first presents the experimental setup used to evaluate \katIALLM and \katIAMA,
followed by the results for our two research questions.
RQ1 evaluates the single prompt refinement variant \katIALLM, while RQ2 evaluates the multi-agent refinement variant \katIAMA.

\subsection{Experimental Setup}
We measured the performance and consistency of \katIALLM considering three different LLMs.
Second, we used the LLM that provided the best balance between performance and result consistency to evaluate \katIAMA.
We also compared the resource usage using this LLM in both variants of \katIA.

Table \ref{tab:llm-selection} lists the three LLMs used to evaluate \katIALLM.
We selected a diverse set of models from three different model families, including both general-purpose and code-focused models.
These API-hosted models are optimized mainly for either inference efficiency or coding tasks while keeping inference costs low.
We accessed all models through OpenRouter\footnote{\url{https://openrouter.ai}} and pinned requests to a single provider per model to improve reproducibility.
Based on token throughput per second, latency, uptime, and cost, we selected \emph{OpenAI} for
GPT-5.4 Mini, \emph{Alibaba Cloud Int.}
for Qwen3 Coder Plus, and \emph{Google Vertex (EU)} for Gemini 2.5 Flash.
This setup makes the evaluation independent of local computing resources.

\begin{table}[htbp]
    \caption{Selected LLMs, their context size, their OpenRouter identifiers, and pinned providers.}
    \begin{center}
        \resizebox{\columnwidth}{!}{%
            \begin{tabular}{lrll}
                \toprule
                \textbf{Model} & \textbf{Context} & \textbf{OpenRouter Identifier} & \textbf{Provider} \\
                \midrule
                GPT-5.4 Mini & 400K & openai/gpt-5.4-mini & OpenAI \\
                Qwen3 Coder Plus & 1M & qwen/qwen3-coder-plus & Alibaba Cloud Int.\\
                Gemini 2.5 Flash & 1M & google/gemini-2.5-flash & Google Vertex (EU)\\
                \bottomrule
            \end{tabular}%
        }
    \end{center}
    \label{tab:llm-selection}
\end{table}

Despite their advantages, LLMs have the fundamental problem of non-determinism, also for code generation~\cite{ouyang2025empirical}.
To assess the consistency of the results while keeping the computational cost manageable, we repeated each experiment for each model and variant three times.
Throughout all experiments, the temperature parameter was set to 0 to maximize output determinism.

Similar to previous approaches, such as \textsc{Ripple}~\cite{Seed2Scope2026Yadavally}, we used set-based evaluation metrics
to measure the performance because \katIA outputs unordered impact sets.
For a commit, we computed precision, recall, and F1-score, based on the ground truth impact sets from our curated dataset and the refined impact set output by \katIA.
Specifically, for each commit and metric, we computed the median across the three repetitions for each model.
We then calculated the macro-averages across all commits for both FQN-matching and CE-matching.
CE-matching only considers the declaring class and the code entity's simple name without any
parameters or further information.
For example, the code entity with the fully qualified name \texttt{org.example.A.b()} gives the class-entity name \texttt{A.b}.
We included CE-matching to allow the comparison with \textsc{Ripple}~\cite{Seed2Scope2026Yadavally}, as they represented their ground truth values as
\texttt{sourceFileSimpleName.methodSimpleName}.
Note that our CE-matching favored \textsc{Ripple}, because we considered class name and entity name, therefore also inner classes, while \textsc{Ripple} only considers top-level classes.
We do not report ranking-based metrics (\eg Hit@k), as our approach predicts an unordered set of impacted entities.

\subsection{Evaluation of \katIALLM}\label{subsec:single-llm-baseline}

\tabref{tab:rq1-model-comparison} lists the results of the three models and initial static impact set as an intermediate step,
evaluated with both FQN-matching and CE-matching.
We can observe that the two models with 1M context size, Gemini 2.5 Flash and Qwen3 Coder Plus
outperform GPT-5.4 Mini for all metrics.
CE-matching consistently outperforms FQN-matching, indicating that the LLMs produce approximate results well but
struggle to accurately identify complete fully qualified names, potentially because they lack access to the source code repository.
For FQN-matching, the single prompt with all three LLMs improves precision compared to the impact set obtained only with static analysis.
GPT-5.4 Mini increases precision by 2.9 percentage points, while the other models more than double the static analysis precision of 15.8\%.
Recall decreases by 3.7 percentage points for GPT-5.4 Mini and by 0.4 percentage points for Gemini 2.5 Flash,
whereas Qwen3 Coder Plus improves the recall by 5.1 percentage points.
Despite these differences, the single prompt approach with the three LLMs achieves higher F1-scores than the static analysis.

For CE-matching, all models substantially improve precision over the static analysis,
with Gemini 2.5 Flash achieving the highest precision of 39.9\%.
Recall also improves for all models except GPT-5.4 Mini, which exhibits a decrease of 2.0 percentage points
compared to the static analysis.
Overall, the single prompt F1-scores for two of the three LLMs are slightly higher than those achieved by \textsc{Ripple}, which reached a precision of 28.2\%, a recall of 36.3\%, and an F1-score of 25.0\%.

\begin{table}[htbp]
    \caption{Macro-averaged precision (Prec), recall (Rec), and F1-score for FQN-matching and CE-matching of the static analysis, \katIALLM across the three models, \katIAMA using Gemini 2.5 Flash, and Ripple$\dagger$.}
    \centering
    \resizebox{\columnwidth}{!}{%
        \begin{tabular}{lrrrrrr}
            \toprule
            & \multicolumn{3}{c}{\textbf{FQN-matching}}
            & \multicolumn{3}{c}{\textbf{CE-matching}} \\
            \cmidrule(lr){2-4}
            \cmidrule(lr){5-7}
            \textbf{Configuration}
            & \textbf{Prec} & \textbf{Rec} & \textbf{F1}
            & \textbf{Prec} & \textbf{Rec} & \textbf{F1} \\
            \midrule

            Static analysis
            & 15.8\% & 23.5\% & 14.6\%
            & 16.9\% & 24.9\% & 15.6\% \\

            \hline

            \addlinespace[2pt]
            \katIALLM (GPT-5.4 Mini)
            & 18.7\% & 19.8\% & 15.8\%
            & 22.1\% & 22.9\% & 18.5\% \\

            \katIALLM (Gemini 2.5 Flash)
            & 31.9\% & 23.1\% & 22.4\%
            & \textbf{39.9\%} & 26.7\% & 27.0\% \\

            \katIALLM (Qwen3 Coder Plus)
            & 32.2\% & 28.6\% & 25.2\%
            & 36.4\% & 32.3\% & 28.5\% \\

            \hline

            \addlinespace[2pt]
            katIA-MA (Gemini 2.5 Flash)
            & \textbf{32.9\%} & \textbf{33.8\%} & \textbf{29.1\%}
            & 37.8\% & \textbf{38.0\%} & \textbf{33.3\%} \\

            \midrule

            Ripple (GPT-4o)\textsuperscript{$^{\dagger}$}
            & -- & -- & --
            & 28.2\% & 36.3\% & 25.0\% \\

            \bottomrule
        \end{tabular}%
    }
    \label{tab:rq1-model-comparison}
\end{table}
\footnotetext{$\dagger$The results of \textsc{Ripple} are taken from the paper of Yadavally and Nguyen~\cite{Seed2Scope2026Yadavally}.
The comparison is an approximation because our CE-matching additionally considers fields and classes, whereas \textsc{Ripple}
predicts only method-level entities.
Furthermore, \textsc{Ripple} uses the source-file name instead of the declaring class name
and reports results for 100 commits, whereas \katIA is evaluated on 263 commits.
A full replication of \textsc{Ripple} was not possible due to missing files in the public repository at the time of our evaluation.}

In addition to their predictive performance, we evaluated the consistency of the results obtained by the three LLMs.
For each sample, we compared the impact sets from the three repetitions, irrespective of their order, and counted the number of
commits for which all three impact sets were identical.
GPT-5.4 Mini was the least consistent model, producing consistent results for only 57 out of the 263 commits.
Qwen3 Coder Plus produced consistent results for 110 out of 263 samples.
Gemini 2.5 Flash achieved the highest consistency with consistent results for 262 commits,
whereas the only inconsistent commit was when the model produced an empty impact set for two of the three repetitions,
while the third repetition's result identified one correct impacted entity and missed the other three.
The consistency analysis shows that LLMs differ not only in predictive performance but also in result consistency,
reinforcing recommendations for empirical studies using LLMs in 
software engineering~\cite{baltes2026guidelinesempiricalstudiessoftware}.

We answer \textbf{RQ1: \RQOne} as follows:
Compared to the static analysis results, all evaluated LLMs produce an impact set with higher precision and F1-scores.
The best performance is achieved by Qwen3 Coder Plus, resulting in
an F1-score of 25.2\% for FQN-matching and 28.5\% for CE-matching.
Two of our three evaluated models outperform \textsc{Ripple} in terms of precision and F1-score.
Qwen3 Coder Plus returned only 41.83\% consistent results,
while Gemini 2.5 Flash achieved comparable results of 22.4\% and 27.0\% F1-score, with 99.62\% consistent results.
GPT-5.4 Mini performed worst with a CE-matching F1-score of 18.5\% and only 21.67\% consistent results.

\subsection{Evaluation of \katIAMA}\label{subsec:multi-agent}

Despite performing best, Qwen3 Coder Plus produced highly inconsistent results.
Therefore, we used only Gemini 2.5 Flash to evaluate \katIAMA.
The respective performance is depicted in the penultimate row in \tabref{tab:rq1-model-comparison}.
Similar to the experiment with \katIALLM, we performed three repetitions and used the median result per commit to compute the performance metrics.
As with \katIALLM, CE-matching achieves better performance across all metrics.
Furthermore, \katIAMA with Gemini 2.5 Flash produced consistent answers across all samples.

For CE-matching, recall and F1-score increase from 26.7\% to 38.0\% and 27.0\% to 33.3\%,
whereas precision decreases from 39.9\% to 37.8\% compared to \katIALLM with Gemini 2.5 Flash.
\katIAMA outperforms \textsc{Ripple} for all metrics in the CE-matching evaluation.
For instance, \katIAMA's F1-score is 33.3\% while \textsc{Ripple} in its best setting achieved an F1-score of 25\%.
Compared to \katIALLM, \katIAMA improves all FQN-matching metrics,
increasing precision from 31.9\% to 32.9\%, recall from 23.1\% to 33.8\%, and F1-score from 22.4\% to 29.1\%.
Overall, we observe that \katIAMA particularly improves the recall while reaching similar or slightly better precision than
the other approaches.

In addition to performance metrics and result consistency, we report token usage, cost, and generation time obtained
from the respective OpenRouter logs.
\tabref{tab:resource-usages} lists the median and quartiles ($[Q_1, Q_3]$) across the 263 commits for \katIALLM and \katIAMA, both using
Gemini 2.5 Flash and the median values among the three repetitions.
In comparison to \katIALLM, \katIAMA increased median token usage by $5.2\times$,
median generation time by $19.7\times$, and costs by $13.0\times$ observed using the pricing options available at the time of the experiments (July 2026).
    It shows that \katIAMA comes with higher token usage, and thus, higher cost and generation time
due to its three model calls instead of \katIALLM's single prompt approach.
Despite that increase, both time and cost are within an acceptable range while improving both performance and result consistency.
\begin{table}[htbp]
        \caption{Resource usage of \katIALLM and \katIAMA with the model Gemini 2.5 Flash.
        Median and the first and third quartiles ($[Q_1,Q_3]$) were extracted across all 263 commits using
        their median over the three respective repetitions.
        All values were extracted from the corresponding OpenRouter logs.}
        \centering
        \footnotesize
        \renewcommand{\arraystretch}{1.15}
        \setlength{\tabcolsep}{1.5pt}
        \begin{tabular*}{\columnwidth}{@{\extracolsep{\fill}}l@{\hspace{6pt}}r@{\,[}r@{,\,}r@{]\hspace{6pt}}r@{\,[}r@{,\,}r@{]\hspace{6pt}}r@{\,[}r@{,\,}r@{]}}
        \toprule
        \textbf{Variant}
        & \multicolumn{3}{c}{\textbf{Generation Time}}
        & \multicolumn{3}{c}{\textbf{Token Usage}}
        & \multicolumn{3}{c@{}}{\textbf{OpenRouter Cost}} \\
        \cmidrule(lr){2-4}\cmidrule(lr){5-7}\cmidrule(lr){8-10}
        & \multicolumn{3}{c}{(s)}
        & \multicolumn{3}{c}{($\times10^3$)}
        & \multicolumn{3}{c@{}}{(US cents)} \\
        \midrule
        \katIALLM
        & 0.570 & 0.444 & 0.738
        & 1.235 & 1.023 & 1.774
        & 0.048 & 0.036 & 0.064 \\
        \addlinespace[3pt]
        \katIAMA
        & 11.214 & 8.298 & 14.301
        & 6.462 & 4.953 & 8.836
        & 0.626 & 0.472 & 0.807 \\
        \bottomrule
        \end{tabular*}
        \label{tab:resource-usages}

\end{table}

With these results, we answer \textbf{RQ2: \RQTwo} as follows:
\katIAMA outperforms the single prompt approach \katIALLM and achieves higher performance metrics than the approach \textsc{Ripple}.
Compared to \katIALLM and using Gemini 2.5 Flash, \katIAMA increases the FQN-matching F1-score from 22.4\% to 29.1\%
and the CE-matching F1-score from 27.0\% to 33.3\%.
Despite the increased performance, \katIAMA needs substantially more resources than \katIALLM.
Nevertheless, these values stay within a reasonable range.
\section{Discussion}\label{sec:discussion}
This section discusses potential implications of the results and threats to the validity of our dataset \dataset, approach \katIA, and evaluation.

\subsection{Implications}

Our results show that the combination of static code analysis and LLMs benefits code change impact analysis.
The consistent increase in precision across the experiments indicates that the LLMs primarily act as filters of the initial static impact set.
The static analysis includes entities connected through program dependencies, even when they are not relevant for the given seed changes.
By additionally considering the seed changes, their specific source code changes, and the corresponding bug report, LLMs can remove candidates that
are structurally related but unlikely to need modification as a consequence of the seed changes.
However, such filtering introduces a trade-off between precision and recall as filtering false positives increases precision, but
it may also wrongly remove impacted entities and thereby reduce the recall.
The results achieved by \katIAMA indicate that performing impact set refinement across three specialized
agents, each responsible for exactly one task, achieves similar precision but higher recall. 
To exactly map the performance and consistency improvements to the agentic setup, an ablation study needs to be performed,
as our results do not report whether the improvement is due to the setup, decomposed prompt, structured intermediate results, or additional model calls.

In addition to predictive performance, our experiments demonstrated that result consistency is an important criterion when evaluating LLMs
for code change impact analysis.
While Qwen3 Coder Plus and Gemini 2.5 Flash both achieved comparably high F1-scores, the results of Qwen3 Coder Plus were less consistent, reducing the practical usability of the approach.

Despite improving the state-of-the-art in code change impact analysis, the absolute performance values indicate that it remains a challenging problem.
Even \katIAMA, which performed best in our experiments, missed a large number of impacted code entities, highlighting the complexity
of code change impact analysis.

\subsection{Threats to Validity}\label{subsec:threats-to-validity}
We discuss threats to external and internal validity,
covering both the generalizability of our findings and potential bias in our study design and execution.

\subsubsection{External Validity}
Our curated dataset \dataset faces several potential threats to external validity.
First, the projects contained in the dataset comprise only open-source Apache Java projects.
Hence, our results might not generalize to other programming languages or projects outside the Apache ecosystem.
This limitation stems from using the set of projects used in \textsc{Alexandria}, a well-established dataset for code change impact analysis.
While we cannot remove this limitation, we improved the data quality of \textsc{Alexandria} by proposing our curated dataset \dataset.
Another threat is that the considered projects are not uniformly represented in the dataset,
as we randomly sampled commits from the projects rather than enforcing an equal number
of commits per project. This imbalance might bias the results towards the projects with a large number of commits.
However, \tabref{tab:dataset} shows that the majority of projects have between 5 and 20 commits, and only one project
has 56 commits.

We randomly sampled 400 commits from the 918 in \textsc{Alexandria} and finally obtained
a dataset that contained 263 manually labeled commits.
This provides a margin of error of 5.11\% within a 95\% confidence level, which we deem acceptable.
We excluded commits without both seed and impacted code entities as keeping them in our dataset might introduce a selection bias.
Similar works also followed this criterion, as they always considered a minimum of two changed methods and then applied their heuristics to determine the
respective seed change.
Furthermore, we considered only commits with one to five changed production code files, which does not cover large commits. 
However, in the original dataset, 824 of 918 commits are within this range, and only 94 are larger commits.
Therefore, our restriction excludes a relatively small proportion of commits which we consider acceptable.

We also did not include files other than Java production code files in our analysis.
We excluded both test files and configuration files.
Consequently, our approach cannot identify impacts outside the production code.
Providing information on Java test files or meta-files, such as XML files, might improve the detection of change impact.

Finally, we performed the experiments with three LLMs and the results might not generalize to all LLMs. 
Many LLMs exist, and the rapid developments reveal new versions frequently.
We mitigated this threat by selecting models from three different model families, including both general-purpose and code-focused models.

\subsubsection{Internal Validity}
One potential threat to internal validity concerns the labeling of the commits. We mitigated this threat
by performing the labeling with three experts and in three phases: a calibration phase, a shared phase, and 
an individual labeling phase. The first two phases were used to evaluate and refine the rules and guidelines then used for labeling the commits in the individual phase.
We considered the informal agreement in the shared phase sufficient to proceed with the individual labeling.
Since there was no dedicated overlap in the individual phase, we did not measure an agreement for this share of data.
    However, the previous labeling phases helped to reduce potential errors.

The prompt is the fundamental part of any LLM call, and therefore represents a potential source of bias.
To mitigate the risk of introducing a prompt bias, we carefully designed all prompts used in this study according to findings and guidelines of recent software engineering research~\cite{sasaki2025landscape, wang2024advanced, baltes2026guidelinesempiricalstudiessoftware}.

In addition, LLMs have many hyperparameters that can lead to different results. 
We tested all LLMs using their default hyperparameter settings and a temperature parameter set to 0, to mitigate this threat.
Furthermore, we addressed LLM non-determinism in our experiments by performing three repetitions per experiment to assess consistency in the results.
We also addressed potential model version drift by providing the model identifier and provider to improve reproducibility.

Another threat to internal validity is the comparability of our work to recent LLM-based code change impact analysis approaches, such as
\textsc{Ripple}~\cite{Seed2Scope2026Yadavally}.
We performed a best-effort comparison by aligning our evaluation with their more coarse-grained matching strategy.
We compared their reported average performance metrics with an approximation of ours,
using simple class and entity names rather than file and entity names. 
This favors \textsc{Ripple} because our entity-name matching also considers inner classes, while the file name used by \textsc{Ripple} does not.
Furthermore, we provide an evaluation based on FQN-matching, which also addresses method overloading.
Finally, \textsc{Ripple} considers only method changes and one seed change, whereas \katIA also considers field and class changes,
newly added methods, and multiple seed changes, which, in our opinion, represents a more realistic setting because not all bugs are
fixed by changing one method.
\section{Conclusion and Future Work}
\label{sec:conclusion}
Code change impact analysis is a fundamental activity in software development. However, this activity is still often performed manually by software developers
because existing approaches struggle with either low precision or low recall. Moreover, appropriate benchmarks are missing to thoroughly evaluate 
change impact analysis approaches.

Addressing the problem with the benchmark, we first presented \dataset, a curated dataset consisting of 263 commits from 22 open-source Java projects
and a total of 1,931 manually labeled changed source code entities. By labeling the changed entities as either seed, impacted, or non-related, we provide a more realistic ground truth for
code change impact analysis, compared to prior work~\cite{EnhancingATHENA2024Yan, Seed2Scope2026Yadavally}.

Next, we presented our approach \katIA that combines static impact analysis with the reasoning capabilities of LLMs. Regarding the integration of LLMs
into our approach, we studied two variants, one using a single prompt called \katIALLM and one using three specialized agents called \katIAMA.
Using \dataset, we evaluated both variants, and, in addition to the classical performance metrics, precision, recall, and F1-score, we also considered
the consistency of the results by repeating each experiment three times.

The results for \katIALLM show that the LLM refines the initial static impact set.
It improves the FQN-matching F1-score from 14.6\% to 22.4\% with Gemini 2.5 Flash and to 25.2\% with Qwen3 Coder Plus when considering the fully qualified name of code entities.
Furthermore, it achieved CE-matching F1-scores 2.0 and 3.5 percentage points higher than the score reported for \textsc{Ripple}~\cite{Seed2Scope2026Yadavally}.
Gemini 2.5 Flash produced consistent results and was used for evaluating \katIAMA.
The results for \katIAMA using Gemini 2.5 Flash showed a further increase
in precision, recall, and F1-score under FQN-matching, with the latter reaching 29.1\%.
Under CE-matching, the F1-score reached 33.3\%.
Overall, the results show that \katIAMA with Gemini 2.5 Flash further improves the performance while keeping inference time and costs low
and that the consistency of predictions should be considered and reported by software engineering studies involving LLMs.

Despite these improvements, code change impact analysis remains a challenging problem. In future work, we plan to
further improve \katIAMA by providing access to the full source code and equipping it with specific tools~\cite{schick2023toolformer}.
We also plan to extend \dataset with more commits from more projects and enrich our labels with change-type information
based on the taxonomy of Fluri \etal~\cite{fluri2007change}.
Furthermore, future work should investigate the factors that caused \katIAMA's performance improvement through ablation studies
and analyze the conversation logs for further information about the contribution of the individual agents and their patterns.

\section*{Acknowledgment}
We used OpenAI Codex to assist with coding, literature search, and polishing of the prompts and the paper.
The authors reviewed and refined the generated material and take full responsibility for the final manuscript.

\bibliographystyle{IEEEtran}
\bibliography{references}

\end{document}